\documentclass[twocolumn,showpacs,preprintnumbers,amsmath,amssymb]{revtex4}
\usepackage{graphicx}
\usepackage{dcolumn}
\usepackage{bm}

\newcommand{\BoldVec}[1]{\mathchoice%
  {\mbox{\boldmath $\displaystyle     #1$}}%
  {\mbox{\boldmath $\textstyle        #1$}}%
  {\mbox{\boldmath $\scriptstyle      #1$}}%
  {\mbox{\boldmath $\scriptscriptstyle#1$}}%
}
\newcommand{\EQ}{\begin{equation}}
\newcommand{\EN}{\end{equation}}
\newcommand{\EQA}{\begin{eqnarray}}
\newcommand{\ENA}{\end{eqnarray}}
\newcommand{\eq}[1]{(\ref{#1})}
\newcommand{\EEq}[1]{Equation~(\ref{#1})}
\newcommand{\Eq}[1]{Eq.~(\ref{#1})}
\newcommand{\Eqs}[2]{Eqs~(\ref{#1}) and~(\ref{#2})}

\newcommand{\App}[1]{Appendix~\ref{#1}}
\newcommand{\Sec}[1]{\S\,\ref{#1}}

\newcommand{\Fig}[1]{Fig.~\ref{#1}}

\newcommand{\Tab}[1]{Table~\ref{#1}}

\newcommand{\bra}[1]{\langle #1\rangle}

\newcommand{\xx}{\BoldVec{x}{}}

\newcommand{\uu}{\BoldVec{u} {}}

\newcommand{\bb}{\BoldVec{b} {}}

\newcommand{\BB}{\BoldVec{B} {}}

\newcommand{\AAA}{\BoldVec{A} {}}

\newcommand{\aaaa}{\BoldVec{a} {}} 

\newcommand{\jj}{\BoldVec{j} {}}
\newcommand{\JJ}{\BoldVec{J} {}}

\newcommand{\EE}{\BoldVec{E} {}}

\newcommand{\nab}{\BoldVec{\nabla} {}}

\newcommand{\oo}{\BoldVec{\omega} {}}

\newcommand{\emf}{\mbox{\boldmath ${\cal E}$} {}}

\newcommand{\dd}{{\rm d} {}}

\def\onethird{{\textstyle{1\over3}}}

\newcommand{\yana}[3]{, Astron. Astrophys. {\bf #2}, #3 (#1).}

\newcommand{\ymn}[3]{, Monthly Notices Roy. Astron. Soc. {\bf #2}, #3 (#1).}

\newcommand{\yjfm}[3]{, J. Fluid Mech. {\bf #2}, #3 (#1).}
\newcommand{\ypr}[3]{, Phys.\ Rev.\ {\bf #2}, #3 (#1).}
\newcommand{\yprl}[3]{, Phys.\ Rev.\ Lett.\ {\bf #2}, #3 (#1).}

\newcommand{\yjgr}[3]{, J. Geophys. Res. {\bf #2}, #3 (#1).}

\newcommand{\yapj}[3]{, Astrophys. J. {\bf #2}, #3 (#1).}

\newcommand{\ypp}[3]{, Phys. Plasmas {\bf #2}, #3 (#1).}

\newcommand{\ypf}[3]{, Phys. Fluids {\bf #2}, #3 (#1).}

\newcommand{\ygafd}[3]{, Geophys. Astrophys. Fluid Dynam. {\bf #2}, #3 (#1).}

\newcommand{\yproc}[4]{, (ed. #3), pp. #2. #4 (#1).}
\newcommand{\ybook}[3]{ {\em #2}. #3 (#1).}

\begin{document}
\preprint{NORDITA 2003-29 AP}

\title{Magnetic helicity evolution in a periodic domain with imposed field}
\author{Axel Brandenburg}
\affiliation{NORDITA, Blegdamsvej 17, DK-2100 Copenhagen \O, Denmark}
\author{William H. Matthaeus}
\affiliation{University of Delaware, Bartol Research Institute, 217 Sharp Lab, Newark, DE 19806}
\date{\today,~ $ $Revision: 1.58 $ $}

\begin{abstract}
In helical hydromagnetic turbulence with an imposed magnetic field
(which is constant in space and time)
the magnetic helicity of the field within a periodic domain
is no longer an invariant of the ideal equations.
Alternatively, there is a generalized magnetic helicity that is an
invariant of the ideal equations.
It is shown that this quantity is
not gauge invariant and that it can therefore not be used in practice.
Instead, the evolution equation of the magnetic helicity of the
field describing the deviation from the imposed field is shown to
be a useful tool.
It is demonstrated that this tool can determine steady state quenching
of the alpha-effect.
A simple three-scale model is derived to describe the
evolution of the magnetic helicity and to predict its sign as
a function of the imposed field strength.
The results of the model agree favorably with simulations.
\end{abstract}
\maketitle

\section{Introduction}

Magnetic helicity has traditionally been used as a diagnostic
tool to characterize magnetic field topology. Only in
recent years magnetic helicity has also become a useful tool in
understanding large scale dynamo action.
Magnetic helicity is important because it is conserved
in the limit of vanishing resistivity.
This is {\it not}\/ the case with the kinetic helicity, which is also
conserved in the inviscid case, but the kinetic helicity dissipation rate
diverges in the inviscid limit \cite{BK01}.
In this sense kinetic helicity is not even approximately conserved at
large Reynolds numbers, while magnetic helicity is very nearly conserved
at large magnetic Reynolds numbers \cite{Ber84}.

A key result that has emerged
from the concept of magnetic helicity conservation is that,
in a periodic domain, a large scale magnetic field generated by
the $\alpha$-effect \cite{Mo78,KR80} saturates on a {\it resistive}\/
time scale \cite{B01}. This time scale can be very long.
The helicity concept has also provided us with
a simple explanation for the final saturation field strength of
helical dynamos in periodic domains; see Ref.~\cite{B01} for details.
In this light, the case with an imposed magnetic field has also been considered
\cite{MMMD02,MMD03}, where it was found that above a certain
field strength the dynamo is suppressed.

In the present paper we use similar ideas to obtain a more detailed
understanding of the case with an imposed field.
We begin with the equation governing the evolution of the
magnetic helicity in a periodic domain in the presence of an imposed
field. It is well known that in that case the magnetic helicity of
the fluctuating magnetic field is no longer conserved in the nonresistive
limit \cite{MG82}, but we also point out that a certain
generalized (or total) magnetic helicity that has sometimes been used
instead is gauge dependent and can therefore not be used in the present
case.
We then discuss applications of the magnetic helicity equation to
the alpha-effect in mean-field electrodynamics and derive a model
equation in order to understand the evolution of the magnetic helicity
for different imposed field strengths.

\section{Magnetic helicity equation}

The evolution of the magnetic field $\BB$ is governed by
\begin{equation}
\frac{\partial\BB}{\partial t}=-\nab\times\EE,
\quad\nab\cdot\BB=0,
\label{dBBdt}
\end{equation}
where the electric field is obtained from Ohm's law,
\begin{equation}
\EE=-\uu\times\BB+\eta\JJ,
\label{ElectricField}
\end{equation}
where $\uu$ is the velocity, $\JJ$ the current density,
and $\eta$ the resistivity.
Throughout this paper we adopt SI units,
but we set the permeability to unity.
We consider all quantities to be triply-periodic
over a cartesian domain.
We consider the case with a finite mean field,
\begin{equation}
\bra{\BB}\equiv\BB_0=\text{const}\neq\bf{0},
\end{equation}
where angular brackets denote full volume averages.
Such averages have no spatial dependence, but they can
still depend on time.
However, because of periodicity, the volume average of the
curl in \Eq{dBBdt} vanishes, and hence $\dd\bra\BB/\dd t=0$.
In other words, $\BB_0$ is not only constant in space,
but it is also constant in time.

Next, we split the field into a mean and a fluctuating component,
$\BB=\BB_0+\bb$,
and introduce the magnetic vector potential for the
fluctuating component via $\bb=\nab\times\aaaa$,
where $\aaaa$ is periodic.
The uncurled induction equation reads
\begin{equation}
\frac{\partial\aaaa}{\partial t}=-(\EE+\nab\phi),
\label{dadt}
\end{equation}
where $\phi$ is the scalar potential.

We now consider the magnetic helicity.
The imposed field is constant in space and does therefore
not contribute to the magnetic helicity.
We therefore consider only the magnetic helicity
of the fluctuating field, $H=\bra{\aaaa\cdot\bb}$.
This quantity is gauge invariant because adding a gradient
term to $\aaaa$ does not change $H$:
\begin{equation}
\bra{(\aaaa+\nab\varphi)\cdot\bb}
=H+\bra{\nab\cdot(\varphi\bb)}-\bra{\varphi\nab\cdot\bb}
=H.
\end{equation}
Here we have used the solenoidality of $\bb$, and the fact that the
volume average over a divergence term vanishes for a periodic domain.
The equation for the (gauge-dependent)
helicity density of the fluctuating
field can be obtained in the form
\begin{equation}
\frac{\partial}{\partial t}(\aaaa\cdot\bb)=-2\EE\cdot\bb
-\nab\cdot[(\EE-\nab\phi)\times\aaaa].
\label{heldensity}
\end{equation}
The divergence term vanishes after volume averaging, so
\begin{equation}
\frac{\dd}{\dd t}\bra{\aaaa\cdot\bb}=-2\bra{\EE\cdot\bb},
\label{heldensityaver}
\end{equation}
where all terms are gauge-independent.
Making use of \Eq{ElectricField}, we have
\begin{equation}
\frac{\dd}{\dd t}\bra{\aaaa\cdot\bb}=
2\bra{(\uu\times\BB_0)\cdot\bb}
-2\eta\bra{\jj\cdot\bb},
\label{dabdt}
\end{equation}
where we have used $\bra{\JJ\cdot\bb}=\bra{\jj\cdot\bb}$.
\EEq{dabdt} can also be written as
\begin{equation}
\frac{\dd}{\dd t}\bra{\aaaa\cdot\bb}=
-2\emf_0\cdot\BB_0-2\eta\bra{\jj\cdot\bb},
\label{helicity_eqn_emf}
\end{equation}
where the electromotive force, $\emf_0=\emf_0(t)=\bra{\uu\times\bb}$,
has been introduced.
If the flow is isotropic and helical, there will be an $\alpha$-effect
\cite{Mo78,KR80,B01} with $\emf_0=\alpha\BB_0$, so
\begin{equation}
\frac{\dd}{\dd t}\bra{\aaaa\cdot\bb}=
-2\alpha\BB_0^2-2\eta\bra{\jj\cdot\bb}.
\label{helicity_eqn}
\end{equation}
(In \App{DisplacementCurrentImposed} we clarify the implications of
a finite $\alpha$ effect when the Faraday displacement current is
restored in the Maxwell equations and when it is ignored.)
Since $\bra{\aaaa\cdot\bb}$ is gauge invariant, it is a physically
meaningful quantity. If there is a steady state, then
$\bra{\aaaa\cdot\bb}$ must also be steady. In that case we have
\begin{equation}
\alpha=-\eta\bra{\jj\cdot\bb}/\BB_0^2,
\label{alpha}
\end{equation}
which is a relation due to Keinigs \cite{Kei83} for the $\alpha$-effect
in the {\it saturated}\/ (steady) state;
see also Ref.~\cite{MGL86}. If the field is weak, the
$\alpha$-effect will remain finite in the high conductivity
limit \cite{KR80}.

The presence of a {\it finite} $\alpha$-effect means that the structure
of \Eq{alpha} is very different when there is an imposed field.
Unlike the case without imposed field ($\BB_0={\bf 0}$), the quantity
$H=\bra{\aaaa\cdot\bb}$ is no longer conserved in the limit $\eta\to0$.
This prompted Matthaeus \& Goldstein \cite{MG82} to consider the
quantity
\begin{equation}
\hat{H}=H+2\AAA_0\cdot\BB_0,
\label{Hhat}
\end{equation}
where
\begin{equation}
\AAA_0(t)=\int_0^t\emf_0(t')\,\dd t'.
\label{AAA0}
\end{equation}
Note that $\hat{H}$ is constructed such that it satisfies the equation
\begin{equation}
\frac{\dd\hat{H}}{\dd t}=-2\eta\bra{\jj\cdot\bb}.
\label{dHhatdt}
\end{equation}
Indeed, this equation reduces to \Eq{helicity_eqn_emf} after inserting
\Eqs{Hhat}{AAA0} into \eq{dHhatdt}.

The symbol $\AAA_0$ is chosen to make the extra term in
\Eq{Hhat} look like a magnetic helicity, even though
$\nabla\times\AAA_0={\bf0}\neq\BB_0$.
The reason for $\nabla\times\AAA_0\neq\BB_0$ is that $\AAA_0$
corresponds to a slowly varying variable, whose curl gives $\BB_0$ at
a {\it higher} order, and not the order we are working in; see Eq.~(A9)
of Ref.~\cite{Stribling_etal1994}.
Conversely, $\BB_0$ corresponds to the curl of $\AAA_0$ at a lower order.
A rigorous scale expansion is given in the Appendix of
Ref.~\cite{Stribling_etal1994}.

In the nonresistive limit, the right hand side of \Eq{dHhatdt}
vanishes, and so $\hat{H}$ is conserved. One may be tempted to
conclude that in the steady state, $\bra{\jj\cdot\bb}=0$. This is not
generally true, however, and it would be in conflict with
\Eqs{helicity_eqn}{alpha}. Certainly for sufficiently
weak fields $\alpha$ is finite \cite{VC92},
so $\eta\bra{\jj\cdot\bb}$ will also remain
finite, see \Eq{alpha}. Therefore, $\hat{H}$ cannot be constant in
the steady state. The reason for this puzzle
is that $\hat{H}$ is not gauge invariant \cite{Ber97}, because
the definition of $\hat{H}$ involves the quantity $\AAA_0$. At first
glance, $\AAA_0$ appears to be gauge invariant,
because $\emf_0$ is gauge invariant
and $\AAA_0$ involves only a time
integral over $\emf_0$; see \Eq{AAA0}. However, the beginning of the time
integration is ill-defined, so in general one can replace
\begin{equation}
\AAA_0(t)\rightarrow
\tilde{\AAA}_0(t)=\AAA_{00}+\int_0^t\emf_0(t')\,\dd t',
\label{add_gauge}
\end{equation}
which would lead to a {\it different}\/ conserved
magnetic helicity, $\hat{H}+\Delta\hat{H}$, where
$\Delta\hat{H}=2\AAA_{00}\cdot\BB_0=\mbox{const}$ is undetermined.
Therefore, $\hat{H}$ is not a physically meaningful quantity,
so it is not surprising that $\hat{H}$ can have a
component that grows linearly in time.
We emphasize however that $H=\bra{\aaaa\cdot\bb}$ is still gauge
invariant and therefore physically meaningful, even though it is no
longer a conserved quantity in the ideal limit.

\section{Non-periodic gauge potentials}

In this section we want to comment on the related issue that adding a
spatially constant vector $\EE_0(t)$ to the right hand side of \Eq{dadt}
would not affect the evolution of $\bb$.
The constant vector $\EE_0(t)$ can readily be absorbed in the definition
of the scalar potential $\phi$, because it is specified only up to an
additional gauge potential.
However, this gauge potential has in general a
non-periodic contribution, even when all other quantities are periodic.
More specifically, $\phi$ in \Eq{dadt}
must have an additional component that varies
{\it linearly}\/ in space, i.e.\
\begin{equation}
\phi=\tilde\phi-\EE_0\cdot\xx,
\label{phi}
\end{equation}
where $\tilde\phi$ is periodic and $\xx$ is the position vector.
We stress that $\phi$ in \Eq{phi} is therefore in general not periodic,
even if $\aaaa$ and $\bb$ are periodic.

A comment on the helicity flux associated with the gauge field is here
in order.
This flux is often written as $\phi\bb$ which would then not be periodic
and hence it is not obvious that its surface integral vanishes; see
\Eqs{heldensity}{heldensityaver}.
(The importance of this term for magnetic helicity injection has been
discussed in Ref.~\cite{JC84}.)
However, using the identity
\begin{equation}
\nab\phi\times\aaaa=\nab\times(\phi\aaaa)-\phi\bb,
\end{equation}
the magnetic helicity flux $\phi\bb$ can also be written as
$\nab\phi\times\aaaa$, which is periodic.
Therefore, there is no contribution from the $\phi\bb$ term in our case.
(We note that similar manipulations can be used to turn a
non-periodic vector potential into a periodic one if the velocity
is a linear function of coordinates \cite{BNST95}.)
The term $\nab\times(\phi\aaaa)$ has recently been discussed in a
formulation of a magnetic helicity conserving dynamo effect \cite{VC01}.
Obviously, such a term does not give a contribution under the
divergence and hence cannot be physically meaningful \cite{AB01}.

\section{Application to $\alpha$-quenching}
\label{Application}

There have been a number of simulations of helically forced periodic
flows with an imposed magnetic field.
The general objective is to obtain the $\alpha$-effect and its suppression
as a function of field strength \cite{B01,TCV93,CH96}.

In the steady state, \Eq{alpha} can be used to determine $\alpha$
by measuring $\bra{\jj\cdot\bb}$ in a simulation with an applied
magnetic field $\BB_0$.
For helical turbulence, $\bra{\jj\cdot\bb}$ can be approximated by
\begin{equation}
\bra{\jj\cdot\bb}\approx\epsilon_{\rm f}k_{\rm f}\bra{\bb^2}, 
\label{jb_eqn}
\end{equation}
where $\epsilon_{\rm f}=\pm1$
for a fully helical field with positive or negative helicity, and
$|\epsilon_{\rm f}|<1$ for fractional helicity.
A strongly helical small scale magnetic field is generally expected when
the turbulent velocity field is also strongly helical \cite{PFL76}.
The steady state $\alpha$ is therefore given by
\begin{equation}
\alpha=-\epsilon_{\rm f}k_{\rm f}\eta\bra{\bb^2}/\BB_0^2
\quad\mbox{(steady state value)}.
\end{equation}
On the other hand, the kinematic value of
$\alpha$ can be estimated in terms of the kinetic helicity,
\begin{equation}
\alpha_{\rm K}=-\onethird\tau\bra{\oo\cdot\uu}
\quad\mbox{(kinematic value)},
\label{alphaK}
\end{equation}
where $\tau$ is the correlation time which, in turn, can be expressed
in terms of the turbulent magnetic diffusivity for which we have a
similar expression, $\eta_{\rm t}=\onethird\tau\bra{\uu^2}$.
In analogy to \Eq{jb_eqn}, we write
$\bra{\oo\cdot\uu}\approx\epsilon_{\rm f}k_{\rm f}\bra{\uu^2}$, so
\begin{equation}
\frac{\alpha}{\alpha_{\rm K}}\approx\frac{\eta}{\eta_{\rm t}}\,
\frac{\bra{\bb^2}}{\BB_0^2}
\quad\mbox{(steady state value)}.
\label{alpha_limiting}
\end{equation}
If we assume that the small scale field is in equipartition, i.e.\
$\bra{\bb^2}\approx\bra{\mu_0\rho\uu^2}\equiv B_{\rm eq}^2$,
and if we define \cite{BB02} the magnetic Reynolds number as
$R_{\rm m}\equiv\eta_{\rm t}/\eta$, then \Eq{alpha_limiting}
can be turned into the interpolation formula,
\begin{equation}
\alpha=\frac{\alpha_{\rm K}}{1+R_{\rm m}\BB_0^2/B_{\rm eq}^2},
\label{catastrophic}
\end{equation}
that recovers \Eq{alpha_limiting} for strong fields and
$\alpha=\alpha_{\rm K}$ in the weak field limit, $\BB_0\to0$.
This equation is known as the catastrophic quenching formula of
Vainshtein and Cattaneo \cite{VC92}.

\begin{figure}[t!]\centering\includegraphics[width=0.5\textwidth]{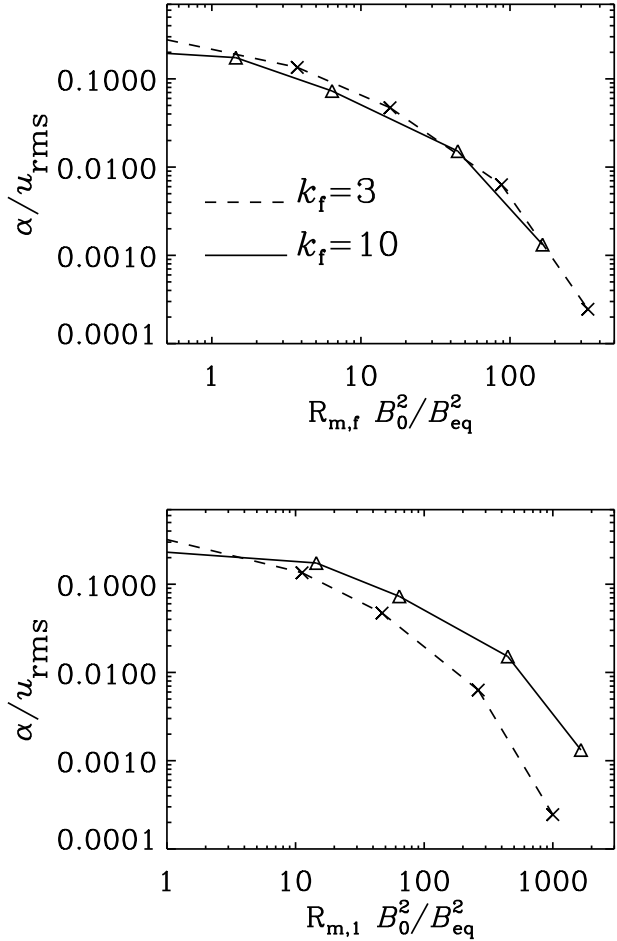}\caption{
Normalized $\alpha$-effect versus normalized magnetic energy, scaled
with the small scale magnetic Reynolds number (upper panel)
and the large scale magnetic Reynolds number (lower panel).
Note that the onset of quenching is governed by the small scale magnetic
Reynolds number, not the large scale magnetic Reynolds number.
}\label{Fpquen}\end{figure}

In order to confirm that the onset of steady state quenching depends on the
magnetic Reynolds number based on the forcing scale \cite{FB02},
$R_{\rm m,f}\equiv u_{\rm rms}/(\eta k_{\rm f})$, and not on the
magnetic Reynolds number based on the scale of the box \cite{B01},
$R_{\rm m,1}\equiv u_{\rm rms}/(\eta k_1)$,
we show two series of simulations obtained for different values of
the forcing wavenumber $k_{\rm f}$, for different values of $B_0$.
The forcing of the flow was fully helical; for details on the
numerical method see Ref.~\cite{B01}.
The result is shown in \Fig{Fpquen}.

\begin{figure}[t!]\centering\includegraphics[width=0.5\textwidth]{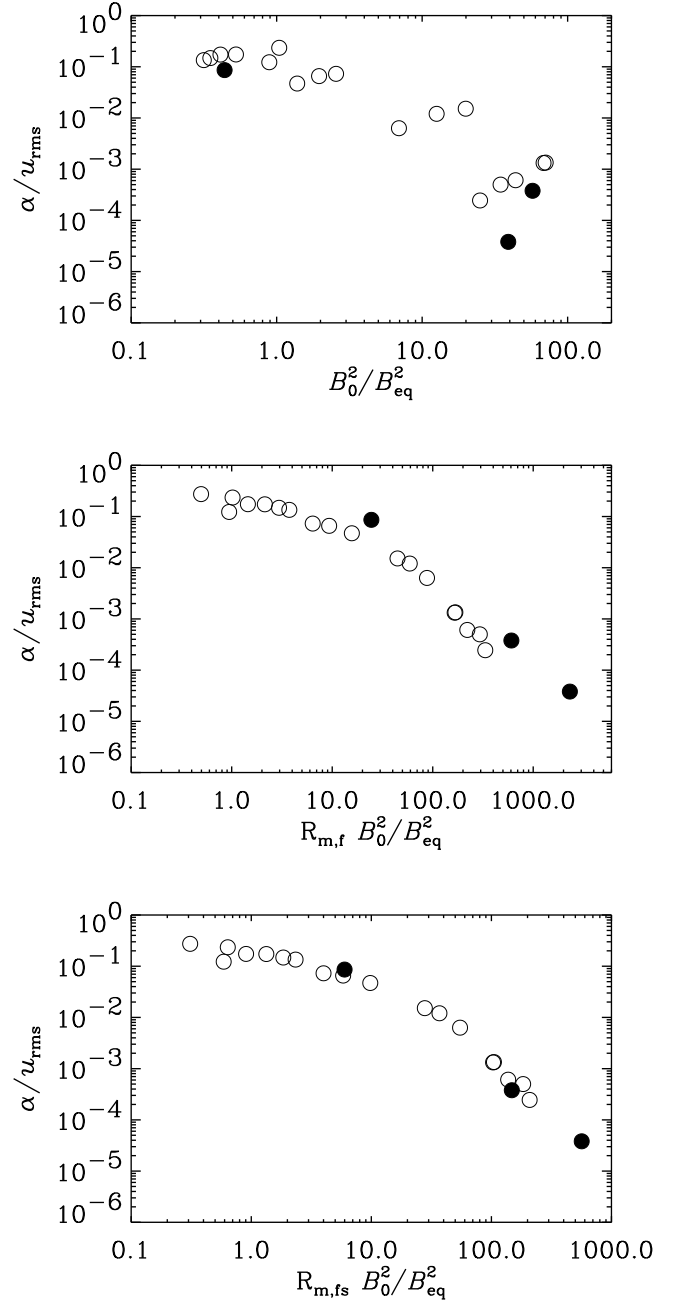}\caption{
Normalized $\alpha$-effect versus magnetic field strength.
The full dots denote runs where hyperdiffusion has been used. In the second
panel, $B_0^2/B_{\rm eq}^2$ is scaled with the magnetic Reynolds number
based on the forcing scale, $R_{\rm m,f}$. The last panel is similar to the
second, but the {\it effective}\/ forcing wavenumber has been used which is
$k_{\rm f}$ scaled by a factor 1.6. This brings especially the hyperdiffusive
runs (full dots) closer to the rest of the data points.
}\label{Fpres}\end{figure}

Next, we consider simulations where we use hyperdiffusivity, i.e.\ the
ordinary magnetic diffusion operator, $\eta\nabla^2\BB$, is replaced
by $(-1)^{n-1}\eta_n\nabla^{2n}\BB$, where $n=1$ corresponds to the standard case.
This is a common tool in order to extend the inertial range of the
turbulence \cite{MFP81}, but it is also clear that this leads to wrong
saturation field strengths \cite{BS02}.
In the presence of hyperdiffusivity, the magnetic Reynolds number is
defined as $R_{\rm m,f}\equiv u_{\rm rms}/(\eta_n k_{\rm f}^{2n-1})$.
In the following we show, however, that the quenching data are better
described by a single quenching curve when $k_{\rm f}$ is rescaled,
\begin{equation}
R_{\rm m,fs}\equiv u_{\rm rms}/(\eta_n k_{\rm fs}^{2n-1}),
\quad\mbox{where $k_{\rm fs}=1.6\times k_{\rm f}$}.
\end{equation}
The result is shown in \Fig{Fpres}.
The factor of 1.6 is not expected to be universal but is probably
a slowly varying function of magnetic Reynolds number \cite{BS02,BB02}.

We conclude that \Eq{catastrophic} describes the simulations quite well
provided the magnetic Reynolds number is defined in a suitable manner.
We emphasize however that this equation only applies to the
steady state and if there is no mean current.
This is generally not the case
and therefore the quenching is in practice not automatically
catastrophic \cite{BB02}, i.e.\
the onset of quenching does not depend on $R_{\rm m}$.

\section{Evolution of large scale magnetic helicity}

Recently, the effect of an imposed
field on the inverse cascade has been studied \cite{MMMD02,MMD03}. If
the imposed magnetic field is weak or absent, there is a strong nonlocal
transfer of magnetic helicity and magnetic energy from the forcing scale
to larger scales. This leads eventually to the accumulation of magnetic
energy at the scale of the box \cite{B01,MFP81,BP99}. As the strength
of the imposed field (wavenumber $k=0$) is increased, the accumulation
of magnetic energy at the scale of the box ($k=1$) becomes more
and more suppressed \cite{MMMD02}.

Qualitatively, this can be understood as the result of two competing
effects: (i) the inverse cascade that produces magnetic helicity of
opposite sign at $k=1$ compared to that at the forcing wavenumber $k_{\rm f}$,
and (ii) the $\alpha$-effect operating on the imposed field producing
magnetic helicity of the same sign at $k=1$ than at $k=k_{\rm f}$.
This is because the sign of the $\alpha$-effect is opposite to the sign of
the magnetic helicity at $k=k_{\rm f}$, and $\alpha$ enters with a minus
sign in the evolution equation \eq{helicity_eqn} of magnetic helicity.
Under the assumption that the turbulence is fully helical,
the critical value $B_*$ of the imposed field can be estimated by
balancing the two terms on the right hand side of
\Eq{helicity_eqn} and by approximating, as in \Sec{Application},
$\alpha\approx\eta_{\rm t}\epsilon_{\rm f}k_{\rm f}$
and $\bra{\jj\cdot\bb}\approx\epsilon_{\rm f}k_{\rm f}B_{\rm eq}^2$.
This yields
\begin{equation}
B_*^2/B_{\rm eq}^2\approx\eta/\eta_{\rm t}=R_{\rm m}^{-1},
\label{critical_Bstar}
\end{equation}
where the last equality is again to be understood as
a definition of the magnetic Reynolds number, see also Ref.~\cite{BB02}.
For $B_0>B_*$ the sign of the magnetic helicity is the same both at $k=1$
and at $k=k_{\rm f}$, while for $B_0<B_*$ the signs are opposite.

\begin{figure}[t!]\centering\includegraphics[width=0.5\textwidth]{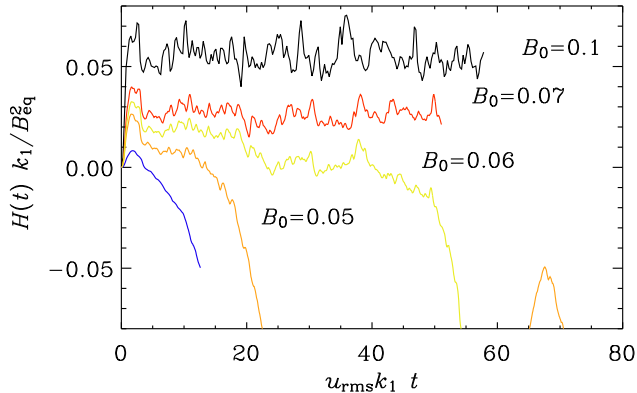}\caption{
Evolution of the total magnetic helicity, $H=H_1+H_{\rm f}$,
as a function of $t$ for different
values of $B_0$, as obtained from the three-dimensional simulation.
}\label{Fphelevol}\end{figure}

A related phenomenological model for saturation of the dynamo effect
under influence of $B_0$ has been given \cite{MMD03}
that is based upon a Fourier scale separation approach. That approach
leads the conclusion that the critical value $B_*\propto R_{\rm m}^{-1}$ rather
than $R_{\rm m}^{-1/2}$ as above. Further analysis may be needed to fully
reconcile the differences in these approaches, both of which appear to have
some support from simulations.

A more quantitative description of the evolution of the magnetic helicity
can be obtained by using a modified two-scale model \cite{FB02,BB02},
where the term $2\emf_0\cdot\BB_0$ from \Eq{helicity_eqn_emf} has been
included, so
\begin{equation}
\dot{H}_1=-2\eta k_1^2 H_1+2\bra{\emf_1\cdot\BB_1}-2\emf_0\cdot\BB_0,
\label{dotH1}
\end{equation}
\begin{equation}
\dot{H}_{\rm f}=-2\eta k_{\rm f}^2 H_{\rm f}-2\bra{\emf_1\cdot\BB_1}.
\label{dotHf}
\end{equation}
Here, $H_1$ and $H_{\rm f}$ are the magnetic helicities at the wavenumbers
$1$ and $k_{\rm f}$, respectively, and $\bra{\emf_1\cdot\BB_1}$ is
the helicity production from $\alpha$-effect and turbulent diffusion
operating on the field at $k=1$.
We note that the sum of \Eqs{dotH1}{dotHf} yields \Eq{heldensityaver}.
The electromotive force $\emf_1$ at wavenumber $k_1$ is given by
\begin{equation}
\emf_1=\alpha_{\rm f}\BB_1-\eta_{\rm t}\JJ_1.
\label{emf1}
\end{equation}
To calculate $\bra{\emf_1\cdot\BB_1}$ in \Eqs{dotH1}{dotHf} we dot
\Eq{emf1} with $\BB_1$, volume average, and note that
$\bra{\JJ_1\cdot\BB_1}=k_1^2H_1$ and $\bra{\BB_1^2}=k_1|H_1|$.
The latter relation assumes that the field at wavenumber $k_1$ is fully
helical, but that it can have either sign.
Thus, we have
\begin{equation}
\bra{\emf_1\cdot\BB_1}=\alpha_{\rm f} k_1|H_1|-\eta_{\rm t}k_1^2 H_1.
\end{equation}
The large scale magnetic
helicity production from the $\alpha$-effect operating on the
imposed field is $\emf_0\cdot\BB_0=\alpha_1\BB_0^2$.
The $\alpha$-effect is proportional to the residual magnetic helicity
of Pouquet, Frisch and L\'eorat \cite{PFL76}, with
\begin{equation}
\alpha=-\onethird\tau
\left(\bra{\oo\cdot\uu}-\bra{\jj\cdot\bb}/\rho_0\right),
\end{equation}
where $\tau$ is the correlation time and $\rho_0$ the average density.
In terms of $H_1$ and $H_{\rm f}$ we write
\begin{equation}
\alpha_1=\alpha_{\rm K}+\onethird\tau k_1^2 H_1,
\end{equation}
\begin{equation}
\alpha_{\rm f}=\alpha_{\rm K}+\onethird\tau k_{\rm f}^2 H_{\rm f},
\end{equation}
for the $\alpha$-effect with feedback from $H_1$ and $H_{\rm f}$,
respectively.
Here, $\alpha_{\rm K}$ is the contribution to the $\alpha$-effect
from the kinematic helicity, as defined in \Eq{alphaK}.

The above set of equations for the case of an imposed magnetic field
is similar to a recently proposed four-scale model \cite{Bla03}, where two
smaller scales were added relative to the two-scale model.
In the present case, on the other hand, instead of including scales
smaller than the forcing scale, the imposed field at the infinite scale
is included, albeit fixed in time.

For finite values of $B_0$, the final value of $H_1$ is particularly
sensitive to the value of $\alpha_{\rm K}$ and turns out to be too
large compared with the simulations.
This disagreement with simulations
is readily removed by taking into account that
$\alpha_{\rm K}=-\onethird\tau\bra{\oo\cdot\uu}$ should itself be quenched
when $B_0$ becomes comparable to $B_{\rm eq}$.
Thus, we write
\begin{equation}
\alpha_{\rm K}=\alpha_{\rm K0}/(1+B_0^2/B_{\rm eq}^2),
\end{equation}
which is a good approximation to more elaborate expressions \cite{RK93}.
We emphasize that this equation only applies to $\alpha_{\rm K}$ and is
therefore distinct from \Eq{catastrophic}.

\begin{figure}[t!]\centering\includegraphics[width=0.5\textwidth]{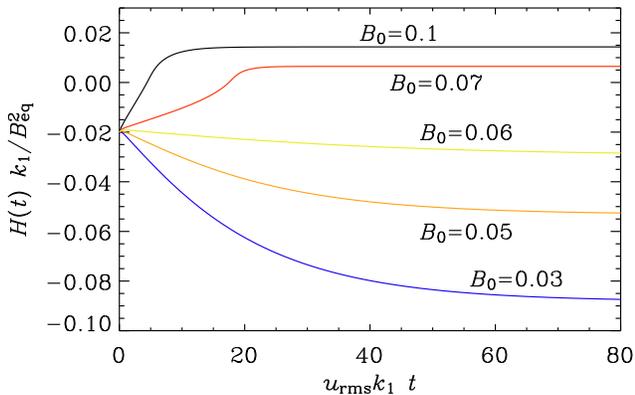}\caption{
Evolution of magnetic helicity as a function of $t$ for different
values of $B_0$, as obtained from the two-scale model.
}\label{Fpall}\end{figure}

In \Fig{Fpall} we show the result of a numerical integration of
\Eqs{dotH1}{dotHf}.
Both the three-dimensional simulation and the two-scale model show a
similar value of $B_0\approx0.06...0.07$, above which $H_1$ changes sign.
This confirms the validity of our estimate of the critical value $B_*$
obtained from \Eq{critical_Bstar}.
Secondly, the time evolution is slow when $B_0<B_*$ and faster when
$B_0>B_*$.
In the simulation, however, the field attains its final level
for $B_0>B_*$ almost instantaneously, which is not the case in the model.
It is possible \cite{MMD03} that the almost instantaneous adjustment in
the simulations is a consequence of the Alfv\'en effect, which is not
included in the present model.
This, and other shortcomings of the present model may also be responsible
for the mismatch between the magnetic helicity amplitudes seen in the
simulations and the model.
Most characteristic in the simulations is the fact that $H_1\to0$ while
$H_{\rm f}\neq0$ in the limit of strong imposed field strength.

\section{Conclusions}

We have shown that (i) in the presence of an imposed field and (ii)
using triple-periodic boundary conditions, the generalized magnetic
helicity \cite{MG82} in \Eq{Hhat} is not gauge-invariant and can
therefore not be used for practical purposes.
This quantity has frequently been used in the solar wind community
as an alternative to the ordinary magnetic helicity which is known
not to be conserved in the limit of vanishing resistivity.
We have argued, however, that even through the ordinary magnetic helicity is
not conserved in the presence of an imposed field and in the limit of
low resistivity, it remains an extremely useful quantity that has
predictive power -- similar to the case without imposed field \cite{B01}.

Based on analytic considerations and confirmed by the simulations,
we have shown that the {\it sign} of the magnetic helicity depends on
the strength of the imposed magnetic field.
If the field is weak enough, the situation is similar to the case
without imposed magnetic field and the sign of the magnetic helicity
is opposite to the sign of the helicity of the turbulence.
If the field exceeds a certain threshold, which is $R_{\rm m}^{-1/2}$
times the equipartition field strength, where $R_{\rm m}$ is the magnetic
Reynolds number based on the forcing wavenumber, the sign of magnetic
helicity changes and becomes equal to the sign of the helicity of the
turbulence.
This can be understood as a consequence of the $\alpha$-effect
operating on the imposed field.
In finite systems, this $\alpha$-effect would cause the large scale
field to have opposite helicity compared to the small scale field.
In an infinite (or periodic) system, this is not possible, and the
entire field in the computational domain plays the role of a small scale field which
must then have the same sign of helicity as the turbulence.

The two-scale model used to describe the nonlinear evolution of helical
dynamos \cite{FB02,BB02} can be generalized to take account of the
large scale field.
The formalism is similar to a recently proposed four-scale model
\cite{Bla03}.
The nonlinear two-scale and multi-scale models play important roles
in modern mean-field dynamo theory.
Given that we are still lacking a proper understanding of solar and
stellar dynamos in the nonlinear regime, an independent confirmation
of the nonlinear multi-scale model must therefore be regarded as a
crucial step toward understanding the origin and maintenance of
magnetic fields in turbulent astrophysical bodies.

\acknowledgments
We thank the referee for inspiring us to comment on the role
of the displacement current in triply periodic systems,
Eric Blackman for comments on the manuscript,
and George Field for organizing the workshop at Virgin Gorda,
where we started working on this paper.
The Danish Center for Scientific Computing is acknowledged for
granting supercomputing time on the Horseshoe machine in Odense.
\\

\appendix

\section{The role of the displacement current in triply periodic systems
with imposed field}
\label{DisplacementCurrentImposed}

In this appendix we discuss a paradoxical situation \cite{MB99} that
arises when comparing volume averages of the present equations (where
periodic boundary conditions are used and a uniform field is imposed) with
the full Maxwell equations (where the displacement current is included).
The displacement current is given by the fourth Maxwell equation,
\begin{equation}
{1\over c^2}\,{\partial\EE\over\partial t}=\nab\times\BB-\JJ,
\label{Displacement}
\end{equation}
where $c$ is the speed of light.
We recall that the permeability has been put to unity.
Applying volume averages, and noting that the volume average of the
curl vanishes, we have
\begin{equation}
{1\over c^2}\,{\partial\bra\EE\over\partial t}=-\bra\JJ.
\label{DisplacementAver}
\end{equation}
On the other hand, the volume average of Ohm's law \eq{ElectricField} yields
\begin{equation}
\bra{\EE}=-\bra{\uu\times\BB}+\eta\bra{\JJ}.
\label{ElectricFieldAver}
\end{equation}
We assume that there is no net flow, i.e.\ $\bra\uu={\bf 0}$.
Therefore, $\bra{\uu\times\BB}=\bra{\uu\times\bb}$.

In helical hydromagnetic turbulence there is an
$\alpha$-effect \cite{Mo78,KR80,B01}, so
$\emf_0\equiv\bra{\uu\times\bb}=\alpha\BB_0\neq{\bf 0}$,
and therefore $\bra{\EE}\neq{\bf 0}$ and, because of
\Eq{DisplacementAver}, $\bra{\JJ}\neq{\bf 0}$.
The latter condition is, of course, inconsistent with our
assumption that $\bra{\JJ}=\bra{\nab\times\BB}={\bf 0}$.
This discrepancy could be particularly important when considering
the contribution of the volume averaged field to the Lorentz force,
$\bra\JJ\times\bra\BB$, which vanishes in the pre-Maxwellian MHD
approximation, but not when the displacement current is retained
\cite{MB99}.

Eliminating $\bra\EE$ from \Eqs{DisplacementAver}{ElectricFieldAver} we have
\begin{equation}
\left(c^2+\eta{\dd\over\dd t}\right)\bra\JJ=\dot\emf_0,
\label{JJvsEMF}
\end{equation}
where the dot on $\emf_0$ denotes time differentiation.
\EEq{JJvsEMF} could be solved for $\bra\JJ$ either in terms of the Green's
function $\exp[-(t-t')c^2/\eta]$ or via series expansion \cite{MB99},
confirming that $\bra{\JJ}\neq{\bf 0}$.
However, here we are interested in the Lorentz force, so we write
\begin{equation}
\left(c^2+\eta{\dd\over\dd t}\right)\left(\bra\JJ\times\bra\BB\right)
=\dot\emf_0\times\BB_0={\bf 0}.
\label{LorentzForce}
\end{equation}
The right hand side of \Eq{LorentzForce} vanishes, because $\bra\BB=\BB_0$
is independent of time and $\dot\emf_0=\dot\alpha\BB_0$
is parallel to $\BB_0$.
The use of the equation $\emf_0=\alpha\BB_0$ ignores random fluctuations
in time about zero.
In that sense, \Eq{LorentzForce} is strictly valid only when the averages
are also taken over time.

\begin{table}[t!]
\centering
\caption{
Comparison of various volume averages in the pre-Maxwellian approximation
and the full Maxwell equations.
The asterisk denotes that this result is strictly valid only when also
a time average is considered.
}
\label{Tsum}
\begin{ruledtabular}
\begin{tabular}{ccc}
                      &pre-Maxwell&  Maxwell \\
\hline
$\bra{\uu\times\bb}$  &  $\neq{\bf 0}$  &  $\neq{\bf 0}$ \\
$\bra{\EE}$           &  $\neq{\bf 0}$  &  $\neq{\bf 0}$ \\
$\bra{\JJ}$           &   $={\bf 0}$    &  $\neq{\bf 0}$ \\
$\bra\JJ\times\bra\BB$&   $={\bf 0}$    &   $={\bf 0}^*$
\end{tabular}
\end{ruledtabular}
\end{table}

The solution to \Eq{LorentzForce} shows that,
if the Lorentz force from the mean field was vanishing initially,
it must vanish at all times.
\Tab{Tsum} summarizes which of the different volume averages discussed
in this section vanish in the pre-Maxwellian MHD approximation and which
quantities remain finite when the full Maxwell equations are used.

The apparent inconsistency is removed by noting that \Eq{Displacement}
does simply not exist in the pre-Maxwellian MHD formulation
and hence cannot be invoked in the discussion.
[The situation is similar to the incompressibility assumption,
$\nab\cdot\uu=0$,
or the anelastic approximation, $\nab\cdot(\rho\uu)=0$, both of
which do not imply $\partial\rho/\partial t=0$.
Indeed, the original continuity equation is no longer used and
has instead been {\it replaced} by $\nab\cdot\uu=0$
or $\nab\cdot(\rho\uu)=0$, respectively.]
Nevertheless, as far as the Lorentz force is concerned, the
neglect of the displacement current is inconsequential,
because it vanishes in either of the two cases; see \Tab{Tsum}.

The mismatch between $\bra{\JJ}={\bf 0}$ in the pre-Maxwellian
approximation and the exact result, $\bra{\JJ}\neq{\bf 0}$, is negligible,
but can be quantified using a rigorous expansion in terms of slowly and
rapidly varying variables \cite{Stribling_etal1994}.
Such an approach also demonstrates quite nicely that the difficulties
introduced into the periodic model by the presence of a nonzero uniform
mean field are due to imposing periodic boundary conditions on the entire
(infinite volume) system.
If instead (see Appendix of Ref.~\cite{Stribling_etal1994}) the turbulence
is assumed to be modeled as {\it locally} homogeneous, in the statistical
sense, and periodicity is employed in a two scale expansion as a local
leading order model, no such problems emerge.

Paradoxical situations arising from the assumption of triple periodicity
are commonly resolved using scale expansion.
Another such example is the famous Jeans swindle \cite{BT87}, where the
assumed zero order equilibrium state does not obey triple periodicity;
see Refs~\cite{Spitzer78,Chavanis02} for a stability analysis using a
proper equilibrium solution.
We emphasize, however, that the problem with the Jeans swindle is
{\it distinct} from the problem with the displacement current discussed here.
The latter is completely resolved by staying fully within the pre-Maxwellian
formulation, while the former is a true mathematical swindle.



\begin{thebibliography}{99}

\bibitem{BK01}
Brandenburg, A., \& Kerr, R. M.\yproc{2001}{358}
{Quantized Vortex Dynamics and Superfluid Turbulence}
{C.F. Barenghi, R.J. Donnelly, W.F. Vinen}
{Lecture Notes in Physics, Vol.\ {\bf571}, Springer Verlag}

\bibitem{Ber84}
M. Berger\ygafd{1984}{30}{79}

\bibitem{Mo78}
H. K. Moffatt\ybook{1978}
{Magnetic Field Generation in Electrically Conducting Fluids}
{Cambridge University Press, Cambridge}

\bibitem{KR80}
F. Krause and K.-H. R\"adler\ybook{1980}
{Mean-Field Magnetohydrodynamics and Dynamo Theory}
{Akademie-Verlag, Berlin; also Pergamon Press, Oxford}

\bibitem{B01}
A. Brandenburg\yapj{2001}{550}{824}

\bibitem{MMMD02}
D. Montgomery, W. H. Matthaeus, L. J. Milano, and
P. Dmitruk\ypp{2002}{9}{1221}

\bibitem{MMD03}
L. J. Milano, W. H. Matthaeus, and P. Dmitruk\ypp{2003}{10}{2287}

\bibitem{MG82}
W. H. Matthaeus and M. L. Goldstein\yjgr{1982}{87}{6011}

\bibitem{Kei83}
R. K. Keinigs\ypf{1983}{26}{2558}

\bibitem{MGL86}
W. H. Matthaeus, M. L. Goldstein, and S. R. Lantz\ypf{1986}{29}{1504}

\bibitem{Stribling_etal1994}
T. Stribling, W. H. Matthaeus, and S. Ghosh\yjgr{1994}{99}{2567}

\bibitem{VC92}
S. I. Vainshtein and F. Cattaneo\yapj{1992}{393}{165}

\bibitem{Ber97}
M. A. Berger\yjgr{1997}{A 102}{2637}

\bibitem{JC84}
T. H. Jensen and M. S. Chu\ypf{1984}{27}{2881}

\bibitem{BNST95}
A. Brandenburg, \AA. Nordlund, R. F. Stein, and
U. Torkelsson\yapj{1995}{446}{741}

\bibitem{VC01}
E. T. Vishniac and J. Cho\yapj{2001}{550}{752}

\bibitem{AB01}
R. Arlt and A. Brandenburg\yana{2001}{380}{359}

\bibitem{TCV93}
L. Tao, C. Cattaneo, S. I. Vainshtein\yproc{1993}{303}
{Solar and Planetary Dynamos}
{M. R. E. Proctor, P. C. Matthews \& A. M. Rucklidge}
{Cambridge University Press}

\bibitem{CH96}
F. Cattaneo and D. W. Hughes\ypr{1996}{E 54}{R4532}

\bibitem{PFL76}
A. Pouquet, U. Frisch and J. L\'eorat\yjfm{1976}{77}{321}

\bibitem{BB02}
E. G. Blackman and A. Brandenburg\yapj{2002}{579}{359}

\bibitem{FB02}
G. B. Field and E. G. Blackman\yapj{2002}{572}{685}

\bibitem{MFP81}
M. Meneguzzi, U. Frisch, U., Pouquet, A.\yprl{1981}{47}{1060}

\bibitem{BS02}
A. Brandenburg and G. S. Sarson\yprl{2002}{88}{055003}

\bibitem{BP99}
D. Balsara and A. Pouquet\ypp{1999}{6}{89}

\bibitem{Bla03}
E. G. Blackman\ymn{2003}{344}{707}

\bibitem{RK93}
G. R\"udiger and L. L. Kitchatinov\yana{1993}{269}{581}

\bibitem{MB99}
D. C. Montgomery and J. W. Bates\ypp{1999}{6}{2727}

\bibitem{BT87}
J. Binney and S. Tremaine\ybook{1987}{Galactic dynamics}
{Princeton University Press, Princeton, New Jersey, pp.\ 287}

\bibitem{Spitzer78}
L. Spitzer, Jr.\ybook{1978}
{Physical processes in the interstellar medium}
{J. Wiley \& Sons, New York, \S13.3}

\bibitem{Chavanis02}
P. H. Chavanis\yana{2002}{381}{340}

\end{thebibliography}
\end{document}